\documentclass[aps,twocolumn,showpacs]{revtex4}

\usepackage{epsf}
\usepackage{bm}

\begin{document}

\title{Spontaneous decay dynamics in atomically doped carbon nanotubes}

\author{I.~V.~Bondarev\footnote{On leave from the Institute for Nuclear Problems at the
Belarusian State University, Bobruiskaya Str.11, 220050 Minsk,
BELARUS} and Ph.~Lambin}

\affiliation{Facult\'{e}s Universitaires Notre-Dame de la Paix, 61
rue de Bruxelles, 5000 Namur, BELGIUM}

\begin{abstract}
We report a strictly non-exponential spontaneous decay dynamics of
an excited two-level atom placed inside or at different distances
outside a carbon nanotube (CN). This is the result of strong
non-Markovian memory effects arising from the rapid variation of
the photonic density of states with frequency near the CN. The
system exhibits vacuum-field Rabi oscillations, a principal
signature of strong atom-vacuum-field coupling, when the atom is
close enough to the nanotube surface and the atomic transition
frequency is in the vicinity of the resonance of the photonic
density of states. Caused by decreasing the atom-field coupling
strength, the non-exponential decay dynamics gives place to the
exponential one if the atom moves away from the CN surface. Thus,
atom-field coupling and the character of the spontaneous decay
dynamics, respectively, may be controlled by changing the distance
between the atom and CN surface by means of a proper preparation
of atomically doped CNs. This opens routes for new challenging
nanophotonics applications of atomically doped CN systems as
various sources of coherent light emitted by dopant atoms.
\end{abstract}


\maketitle

\section{Introduction}

It has long been recognized that the spontaneous emission rate of
an excited atom is not an immutable property, but that it can be
modified by the atomic environment.~Generally called the Purcell
effect~\cite{Purcell}, the phenomenon is qualitatively explained
by the fact that the local environment modifies the strength and
distribution of the \emph{vacuum} electromagnetic modes with which
the atom can interact, resulting indirectly in the alteration of
atomic spontaneous emission properties.

The Purcell effect took on special significance recently in view
of rapid progress in physics of nanostructures. Here, the control
of spontaneous emission has been predicted to have a lot of useful
applications, ranging from the improvement of existing devices
(lasers, light emitting diodes) to such nontrivial functions as
the emission of nonclassical states of light~\cite{Weisbuch}.~In
particular, the enhancement of the spontaneous emission rate can
be the first step towards the realization of a thresholdless
laser~\cite{Pelton} or a single photon source~\cite{Gerard}.~The
possibility to control atomic spontaneous emission was shown
theoretically for microcavities and
microspheres~\cite{Dung,Welsch,Kimble}, optical
fibers~\cite{Klimov}, photonic crystals~\cite{John}, semiconductor
quantum dots~\cite{Sugawara}. Recent technological progress in the
fabrication of low-dimensional nanostructures has enabled the
experimental investigation of spontaneous emission for
microcavities~\cite{Schniepp}, photonic crystals~\cite{Petrov},
quantum dots~\cite{Gayral}.

Carbon nanotubes (CNs) are graphene sheets rolled-up into
cylinders of approximately one nanometer in diameters.~Extensive
work carried out worldwide in recent years has revealed the
intriguing physical properties of these novel molecular scale
wires~\cite{Dresselhaus,Dai}.~Nanotubes have been shown to be
useful for miniaturized electronic, mechanical, electromechanical,
chemical and scanning probe devices and materials for macroscopic
composites~\cite{Baughman}.~Important is that their intrinsic
properties may be substantially modified in a controllable way by
doping with extrinsic impurity atoms, molecules and
compounds~\cite{Duclaux}.~This opens routes for \emph{new}
challenging nanophotonics applications of atomically doped CN
systems as various sources of coherent light emitted by dopant
atoms. Recent successful experiments on encapsulation of single
atoms into single-wall carbon nanotubes~\cite{Jeong} and their
intercalation into single-wall CN bundles~\cite{Duclaux,Shimoda}
stimulate an analysis of atomic spontaneous emission in such
systems as a first step towards their nanophotonics applications.

Typically, there may be two qualitatively different regimes of
interaction of an atomic excited state with a vacuum
electromagnetic field in the vicinity of the~CN. They are the weak
coupling regime and the strong coupling regime~\cite{Eberly}.~The
former is characterized by the monotonous exponential decay
dynamics of the upper atomic state with the decay rate altered
compared with the free-space value.~The latter is, in contrast,
strictly non-exponential and~is characterized by reversible Rabi
oscillations where the energy of the initially excited atom is
periodically exchanged between the atom and the~field. In the
present paper, we develop the quantum theory of the spontaneous
decay of an excited two-level atom near a CN and derive the
evolution equation of the upper state of the system. By solving~it
numerically, we demonstrate the strictly non-exponential
spontaneous decay dynamics in the case where the atom is close
enough to the CN surface.~In certain cases, the system exhibits
vacuum-field Rabi oscillations --- a result already detected for
quasi-2D excitonic and intersubband electronic transitions in
semiconductor quantum microcavities~\cite{Weisbuch1,Sorba} and
never reported for atomically doped CNs so far.

The rest of the paper is arranged as follows. Section~\ref{model}
presents a theoretical model we use to derive the evolution
equation of the upper state population probability amplitude of
the composed quantum system "a~two-level atom interacting with a
quantized CN-modified vacuum radiation field".~In describing
atom--field interaction we follow the original line of
Refs.~\cite{Dung,Welsch}, adopting their electromagnetic field
quantization scheme for a~particular case of the field near an
infinitely long achiral single-wall CN. The carbon nanotube is
considered as an infinitely thin anisotropically conducting
cylinder.~Its surface conductivity is represented in terms of the
$\pi$-electron dispersion law obtained in the tight-binding
approximation with allowance made for azimuthal electron momentum
quantization and axial electron momentum
relaxation~\cite{Slepyan}.~Only the axial conductivity is taken
into account and the azimuthal one, being strongly suppressed by
transverse depolarization fields~\cite{Tasaki,Jorio,Li,Marinop},
is neglected.~In Section~\ref{qualitatively}, the time dynamics of
the upper state population probability is analyzed qualitatively
in terms of two different approximations admitting analytical
solutions of the evolution equation derived in
Section~\ref{model}.~They are the Markovian approximation and the
single-resonance approximation of the density of photonic states
in the vicinity of the CN. Section~\ref{numerically} presents and
discusses the results of the numerical solution of the evolution
equation for various particular cases where the atom is placed in
the center and near the wall inside, and at different distances
outside achiral CNs of different radii.~A summary and conclusions
of the work are given in Section~\ref{conclusion}.

\begin{figure}[t]
\epsfxsize=8.65cm\centering{\epsfbox{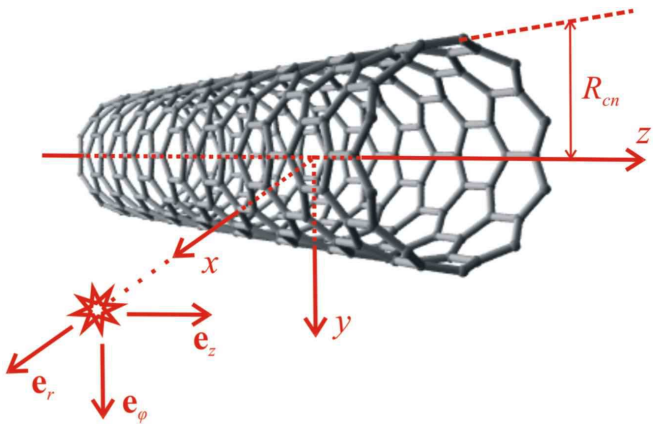}} \caption{(Color
online) The geometry of the problem.} \label{fig1}
\end{figure}

\section{The model}\label{model}

The quantum theory of the spontaneous decay of excited atomic
states in the presence of the CN involves an electromagnetic field
quantization procedure.~Such~a procedure faces difficulties
similar to those in quantum optics of 3D Kramers-Kronig dielectric
media where the canonical quantization scheme commonly used does
not work since, because of absorption, corresponding operator
Maxwell equations become non-Hermitian~\cite{Vogel}.~As a result,
their solutions cannot be expanded in power orthogonal modes and
the concept of modes itself becomes more subtle.~We, therefore,
use an alternative quantization scheme developed in
Refs.~\cite{Dung,Welsch}, where the Fourier-images of electric and
magnetic fields are considered as quantum mechanical observables
of corresponding electric and magnetic field operators.~The latter
ones satisfy the Fourier-domain operator Maxwell equations
modified by the presence of a so-called operator noise current
density $\underline{\hat{\mathbf{J}}\!}\,(\mathbf{r},\omega)$
written in terms of a 3D vector bosonic field operator
$\hat\mathbf{f}(\mathbf{r},\omega)$ and a medium dielectric tensor
$\bm\epsilon(\mathbf{r},\omega)$ (supposed to be diagonal) as
\begin{equation}
{\underline{\hat{J}\!}_{\,i}}(\mathbf{r},\omega)={\omega\over{2\pi}}
\sqrt{\hbar\,\mbox{Im}\epsilon_{ii}(\mathbf{r},\omega)}\,
\hat{f}_{i}(\mathbf{r},\omega),\hskip0.1cm i=1,2,3.
\label{current}
\end{equation}
This operator is responsible for correct commutation relations of
the electric and magnetic field operators in the presence of
medium-induced absorbtion.~In this formalism, the electric and
magnetic field operators are expressed in terms of a continuum
set~of the 3D vector bosonic fields
$\hat{\mathbf{f}}(\mathbf{r},\omega)$ by means of the convolution
over~$\mathbf{r}$ of the current (\ref{current}) with the
\emph{classical} electromagnetic field Green tensor of the
system.~The bosonic field operators
$\hat{\mathbf{f}}^{\dag}(\mathbf{r},\omega)$ and
$\hat{\mathbf{f}}(\mathbf{r},\omega)$ create and annihilate
single-quantum electromagnetic medium excitations.~They are
defined by their commutation relations and play the role of the
fundamental dynamical variables in terms of which the Hamiltonian
of the composed system "electromagnetic field + dissipative
medium" is written in a standard secondly quantized form as
\begin{equation}
\hat{H}=\int\!d\mathbf{r}\!\int_{0}^{\infty}\!\!\!d\omega\,
\hbar\omega\,\hat{\mathbf{f}}^{\dag}(\mathbf{r},\omega)\!
\cdot\!\hat{\mathbf{f}}(\mathbf{r},\omega). \label{H_field}
\end{equation}

Consider a two-level atom positioned at an arbitrary point
$\mathbf{r}_{A}$ near the CN.~Since the problem has a cylindric
symmetry, it is convenient to assign the orthonormal cylindric
basis $\{\mathbf{e}_{r},\mathbf{e}_{\varphi},\mathbf{e}_{z}\}$ in
such a way that $\mathbf{r}_{A}=r_{A}\mathbf{e}_{r}=\{r_{A},0,0\}$
and $\mathbf{e}_{z}$ is directed along the nanotube axis (see
Figure~\ref{fig1}).~Let the atom interact with a quantized
electromagnetic field via an electric dipole transition of
frequency $\omega_{A}$.~The atomic dipole moment may be assumed to
be directed along the CN axis,
$\mathbf{d}=d_{z}\textbf{e}_{z}$.~The contribution of the
transverse dipole moment orientation is suppressed because of
strong depolarization of the transverse field in an isolated
CN~\cite{Tasaki,Jorio,Li,Marinop}.~Strong transverse
depolarization along with transverse electron momentum
quantization allow one to neglect the azimuthal current and radial
polarizability~\cite{Slepyan}, in which case the dielectric tensor
components $\epsilon_{rr}$ and $\epsilon_{\varphi\varphi}$ are
identically equal to unit.~The component $\epsilon_{zz}$ is caused
by the CN longitudinal polarizability and is responsible for the
axial surface current parallel to the $\textbf{e}_{z}$
vector.~This current may be represented in terms of the 1D bosonic
field operators by analogy with Eq.~(\ref{current}).~Indeed,
taking into account the dimensionality conservation in passing
from bulk to a~monolayer in Eq.~(\ref{H_field}) and using a simple
Drude relation~\cite{Tasaki}
\begin{equation}
\sigma_{zz}(\mathbf{R},\omega)=-i\omega{\epsilon_{zz}(\mathbf{R},\omega)-
1\over{4\pi S\rho_{T}}}, \label{sigmaCN}
\end{equation}
where $\mathbf{R}=\!\{R_{cn},\phi,Z\}$~is the radius-vector of an
arbitrary point of the CN surface, $R_{cn}$ is the radius of the
CN, $\sigma_{zz}(\mathbf{R},\omega)$ is the CN surface axial
conductivity per unit length, $S$ is the area of a single
nanotube, $\rho_{T}$ is the tubule density in a bundle, one
immediately has from Eq.~(\ref{current})
\begin{equation}
\underline{\hat{\mathbf{J}}\!}\,(\mathbf{R},\omega)\!=
\!\sqrt{\hbar\omega\mbox{Re}\sigma_{zz}(\mathbf{R},\omega)\over{\pi}}\,
\hat{f}(\mathbf{R},\omega)\textbf{e}_{z} \label{currentCN}
\end{equation}
with $\hat{f}(\mathbf{R},\omega)$ being the 1D bosonic field
operator defined on the CN surface. The total Hamiltonian of the
system under consideration is then written in terms of
$\hat{f}^{\dag}(\mathbf{R},\omega)$ and
$\hat{f}(\mathbf{R},\omega)$ operators as
\[
\hat{{\cal{H}}}=\int\!d\mathbf{R}\!\int_{0}^{\infty}\!\!\!d\omega\,
\hbar\omega\,\hat{f}^{\dag}(\mathbf{R},\omega)\hat{f}(\mathbf{R},\omega)
+{1\over{2}}\,\hbar\omega_{A}\,\hat{\sigma}_{z}
\]\vspace{-0.5cm}
\begin{equation}
-[\,\hat{\sigma}^{\dag}\,\hat{E}^{(+)}_{z}(\mathbf{r}_{A})\,d_{z}+\mbox{h.c.}\,]\,.
\label{Ham}
\end{equation}
Here, the three terms represent the electromagnetic field modified
by the presence of the CN, the two-level atom and their
interaction (within the framework of electric dipole, and rotating
wave approximations~\cite{Dung}), respectively.~The Pauli
operators $\hat{\sigma}_{z}\!=\!|u\rangle\langle
u|-|l\rangle\langle l|$, $\hat{\sigma}\!=\!|l\rangle\langle u|$,
$\hat{\sigma}^{\dag}\!=\!|u\rangle\langle l|$ describe the
two-level atomic subsystem where $|u\rangle$ and $|l\rangle$ are
the upper and lower atomic states, respectively.~The operators
$\hat{\mathbf{E}}^{(\pm)}(\mathbf{r}_{A})$ represent the electric
field the atom interacts with.~For an arbitrary
$\mathbf{r}\!=\!\{r,\varphi,z\}$, they are defined as follows
\[
\hat{\mathbf{E}}(\mathbf{r})=\hat{\mathbf{E}}^{(+)}(\mathbf{r})+
\hat{\mathbf{E}}^{(-)}(\mathbf{r}),
\]\vspace{-0.5cm}
\begin{equation}
\hat{\mathbf{E}}^{(+)}(\mathbf{r})=\int_{0}^{\infty}
\underline{\hat{\mathbf{E}}}(\mathbf{r},\omega)\;d\omega\,,\;\;\;
\hat{\mathbf{E}}^{(-)}(\mathbf{r})=[\hat{\mathbf{E}}^{(+)}(\mathbf{r})]^{\dag}.
\label{Ew}
\end{equation}
Here, $\underline{\hat{\mathbf{E}}}(\mathbf{r},\omega)$ satisfies
the Fourier-domain Maxwell equations
\[
\nabla\times\underline{\hat{\mathbf{E}}}(\mathbf{r},\omega)=
ik\,\underline{\hat{\mathbf{H}}}(\mathbf{r},\omega),
\]\vspace{-0.5cm}
\begin{equation}
\nabla\times\underline{\hat{\mathbf{H}}}(\mathbf{r},\omega)=
-ik\,\underline{\hat{\mathbf{E}}}(\mathbf{r},\omega)+
{4\pi\over{c}}\underline{\hat{\mathbf{I}}}(\mathbf{r},\omega),\label{Maxwell}
\end{equation}
where $\underline{\hat{\mathbf{H}}}(\mathbf{r},\omega)$ stands for
the magnetic field operator [defined by analogy with
Eq.~(\ref{Ew})], $k=\omega/c$, and
\begin{equation}
\underline{\hat{\mathbf{I}}}(\mathbf{r},\omega)\!=\!\!\!\int\!\!d\mathbf{R}\,
\delta(\mathbf{r}-\mathbf{R})\,\underline{\hat{\mathbf{J}}\!}\,(\mathbf{R},\omega)\!
=\!2\underline{\hat{\mathbf{J}}\!}\,(R_{cn},\varphi,z,\omega)\delta(r-R_{cn})
\label{Irw}
\end{equation}
is the exterior operator current density [with
$\underline{\hat{\mathbf{J}}\!}\,(\mathbf{R},\omega)$ defined by
Eq.~(\ref{currentCN})] associated with the presence of the CN.

From Eqs.~(\ref{Maxwell}) and (\ref{Irw}) in view of
Eq.~(\ref{currentCN}), it follows that
\begin{equation}
\underline{\hat{\mathbf{E}}}(\mathbf{r},\omega)=
i{4\pi\over{c}}\,k\!\int\!\!d\mathbf{R}\,\mathbf{G}(\mathbf{r},\mathbf{R},\omega)
\!\cdot\!\underline{\hat{\mathbf{J}}\!}\,(\mathbf{R},\omega)
\label{Erw}
\end{equation}
[and $\underline{\hat{\mathbf{H}}}=(ik)^{-1}
\nabla\times\underline{\hat{\mathbf{E}}}$ accordingly], where
$\mathbf{G}(\mathbf{r},\mathbf{R},\omega)$ is the Green tensor of
the \emph{classical} electromagnetic field in the vicinity of the
CN. Its components satisfy the equation
\begin{equation}
\sum_{\alpha=r,\varphi,z}\!\!\!
\left(\mathbf{\nabla}\!\times\mathbf{\nabla}\!\times-\,k^{2}\right)_{\!z\alpha}
G_{\alpha z}(\mathbf{r},\mathbf{R},\omega)=
\delta(\mathbf{r}-\mathbf{R}), \label{GreenequCN}
\end{equation}
together with the radiation conditions at infinity and the
boundary conditions on the CN surface.~The Hamiltonian~(\ref{Ham})
along with Eqs.~(\ref{Ew})--(\ref{GreenequCN}) and
(\ref{currentCN}) is the modification of the Jaynes--Cummings
model~\cite{Vogel} for an atom in the vicinity of a~solitary~CN.
The classical electromagnetic field Green tensor of this system is
derived in Appendix~\ref{appA}.

When the atom is initially in the upper state and the field
subsystem is in vacuum, the time-dependent wave function of the
whole system can be written as
\[
|\psi(t)\rangle=C_{u}(t)\,e^{-i(\omega_{A}/2)t}|u\rangle|\{0\}\rangle
\]\vspace{-0.5cm}
\begin{equation}
+\int\!d\mathbf{r}\!\int_{0}^{\infty}\!\!\!d\omega\,
C_{l}(\mathbf{r},\omega,t)\,e^{-i(\omega-\omega_{A}/2)t}
|l\rangle|\{1(\mathbf{r},\omega)\}\rangle, \label{wfunc}
\end{equation}
where $|\{0\}\rangle$ is the vacuum state of the field subsystem,
$|\{1(\mathbf{r},\omega)\}\rangle$ is its excited state where the
field is in a single-quantum Fock state, $C_{u}$ and $C_{l}$ are
the population probability amplitudes of the upper state and lower
state of the \emph{whole} system, respectively.~In view of
Eqs.~(\ref{Ew}),~(\ref{Erw}),~(\ref{currentCN}) and the integral
relationship
\[
\mbox{Im}\,G_{\alpha\beta}(\mathbf{r},\mathbf{r}^{\prime},\omega)=
\]\vspace{-0.5cm}
\[
{4\pi\over{c}}\,k\!\int\!d\mathbf{R}\,\mbox{Re}\sigma_{zz}(\mathbf{R},\omega)
G_{\alpha z}(\mathbf{r},\mathbf{R},\omega)G^{\ast}_{\beta
z}(\mathbf{r}^{\prime},\mathbf{R},\omega)
\]
(which is nothing but a particular case of the general 3D Green
tensor integral relationship rigorously proven in
Ref.~\cite{Welsch} with Eq.~(\ref{sigmaCN}) taken into account),
the Schr\"{o}dinger equation with the Hamiltonian~(\ref{Ham})~and
wave function~(\ref{wfunc}) yields the following evolution law for
the population probability amplitude of the upper state of the
system
\begin{equation}
C_{u}(\tau)=1+\int_{0}^{\tau}\!K(\tau-\tau^{\prime})\,
C_{u}(\tau^{\prime})\,d\tau^{\prime}, \label{Volterra}
\end{equation}
\begin{equation}
K(\tau-\tau^{\prime})={\hbar\Gamma_{0}(x_{A})\over{4\pi
x_{A}^{3}\gamma_{0}}}\!\int_{0}^{\infty}\!\!\!\!\!dx\,x^{3}\xi(x)
{e^{-i(x-x_{A})(\tau-\tau^{\prime})}-1\over{i\,(x-x_{A})}}.
\label{kernel}
\end{equation}
Here,
\begin{equation}
x={\hbar\omega\over{2\gamma_{0}}}\hskip0.5cm\mbox{and}\hskip0.5cm
\tau={2\gamma_{0}t\over{\hbar}} \label{dimless}
\end{equation}
are the dimensionless frequency and time, respectively, with
$\gamma_{0}=2.7$~eV being the carbon nearest neighbor hopping
integral~\cite{Wallace} appearing in the CN surface axial
conductivity~in~Eq.~(\ref{currentCN}), $\Gamma_{0}$ is the rate of
the \emph{exponential} atomic spontaneous decay in vacuum obtained
from the general expression of the form~\cite{Barnett,Agarwal75}
\begin{equation}
\Gamma(x)={8\pi d_{z}^{2}\over{\hbar c^{2}}}
\left({2\gamma_{0}x\over{\hbar}}\right)^{\!2}
\mbox{Im}\,G_{zz}(\mathbf{r}_{A},\mathbf{r}_{A},x) \label{Gamma}
\end{equation}
by substituting
\begin{equation}
\mbox{Im}\,G^{v}_{zz}(\mathbf{r}_{A},\mathbf{r}_{A},x)={1\over{6\pi
c}}{2\gamma_{0}x\over{\hbar}}
\label{ImGreenvac}
\end{equation}
for the vacuum imaginary Green tensor~\cite{Abrikosov}.~The
function $\xi(x)$ is the relative density (with respect to vacuum)
of photonic states near the CN given for $r_{A}\!>\!R_{cn}$ (see
Appendix~\ref{appB}) by
\begin{equation}
\xi(x)={\Gamma(x)\over{\Gamma_{0}(x)}}=1+{3\over{\pi}}\,\mbox{Im}\!\!\!
\sum_{p=-\infty}^{\infty} \label{ksi}
\end{equation}\vspace{-0.5cm}
\[
\int_{C}\!{dy\,s(R_{cn},x)v(y)^{4}I_{p}^{2}[v(y)u(R_{cn})x]K_{p}^{2}[v(y)u(r_{A})x]
\over{1+s(R_{cn},x)v(y)^{2}I_{p}[v(y)u(R_{cn})x]K_{p}[v(y)u(R_{cn})x]}}\,,
\]
where $I_{p}$ and $K_{p}$ are the modified cylindric Bessel
functions, $v(y)\!=\!\sqrt{y^{2}-1}\,$,
$u(r)\!=\!2\gamma_{0}r/\hbar c$, and $s(R_{cn},x)\!=\!2i\alpha
u(R_{cn})x\overline{\sigma}_{zz}(R_{cn},x)$ with
$\overline{\sigma}_{zz}\!=\!2\pi\hbar\sigma_{zz}/e^{2}$ being the
dimensionless conductivity and $\alpha\!=\!e^{2}/\hbar
c\!=\!1/137$ representing the fine-structure constant. The
integration contour $C$ goes along the real axis of the complex
plane and envelopes the branch points $y\!=\!\pm 1$ of the
function $v(y)$ in the integrand from below and from above,
respectively.~For $r_{A}\!<\!R_{cn}$, Eq.~(\ref{ksi}) is modified
by the simple replacement $r_{A}\!\leftrightarrow\!R_{cn}$ in the
Bessel function arguments in the numerator of the integrand.~Note
the divergence of $\xi(x)$ at $r_{A}\!=\!R_{cn}\,$, i.~e. when the
atom is located right on the CN surface.~The point is that the~CN
dielectric tensor longitudinal component $\epsilon_{zz}$ [which,
according to Eq.~(\ref{sigmaCN}), is responsible for the surface
axial conductivity $\sigma_{zz}$ in Eq.~(\ref{ksi})] is obtained
as a result of a standard procedure of \emph{physical} averaging
a~local electromagnetic field over the two spatial directions in
the graphene plane~\cite{Lambin}.~Such averaging does not assume
extrinsic atoms on the graphene surface. To take them into
consideration the averaging procedure must be modified. Thus, the
applicability domain of our model is restricted by the condition
\begin{equation}
\mid\!r_{A}\!-R_{cn}\!\mid\,>a\,, \label{condition}
\end{equation}
where $a=1.42$~\AA~ is a graphene interatomic
distance~\cite{Wallace}.

Eq.~(\ref{Volterra}) is a well-known Volterra integral equation of
the second kind.~In our case, it describes the spontaneous decay
dynamics of the excited two-level atom in the vicinity of the
CN.~All the CN parameters that are relevant for the spontaneous
decay are contained in the relative density of photonic states
(\ref{ksi}) appearing in the kernel~(\ref{kernel}).~The density of
photonic states is, in turn, determined by the imaginary classical
Green tensor of~the CN-modified electromagnetic field via
$\Gamma(x)$ given by Eq.~(\ref{Gamma}).

\section{Qualitative analysis}\label{qualitatively}

In this Section we will qualitatively analyze the time dynamics of
the upper state population probability $C_{u}(\tau)$ in terms of
two different approximations admitting analytical solutions of the
evolution problem~(\ref{Volterra}),~(\ref{kernel}). They are the
Markovian approximation and the single-resonance approximation of
the relative density $\xi(x)$ of photonic states in the vicinity
of the CN.

\subsubsection{Markovian approximation}

In the case where the Markovian approximation is applicable, or,
in other words, when the atom-field coupling strength is weak
enough for atomic motion memory effects to be insignificant, so
that they may be neglected, the time-dependent factor in the
kernel~(\ref{kernel}) may be replaced by its long-time
approximation
\[
{e^{-i(x-x_{A})(\tau-\tau{^\prime})}-1\over{i(x-x_{A})}}\rightarrow
-\pi\delta(x-x_{A})+i{\cal{P}}{1\over{x-x_{A}}}
\]
(${\cal{P}}$ denotes a principal value). Then, in view of
Eq.~(\ref{ksi}), the kernel becomes
\[
K(\tau-\tau{^\prime})=-{\hbar\Gamma(x_{A})\over{4\gamma_{0}}}+i\Delta(x_{A})
\]
with
\[
\Delta(x_{A})={\hbar\Gamma_{0}(x_{A})\over{4\gamma_{0}\pi
x_{A}^{3}}}\,{\cal{P}}\!\int_{0}^{\infty}\!\!\!dx\,x^{3}{\xi(x)\over{x-x_{A}}},
\]
and Eq.~(\ref{Volterra}) yields
\begin{equation}
C_{u}(\tau)=\exp\left\{\!\left[-{\hbar\Gamma(x_{A})\over{4\gamma_{0}}}
+i\Delta(x_{A})\right]\!\tau\!\right\}
\label{Cumark}
\end{equation}
--- the exponential decay dynamics of the [shifted by
$\Delta(x_{A})$] upper atomic level with the rate
$\Gamma(x_{A})$.~This case was analyzed in Ref.~\cite{Bondarev02}.

\subsubsection{Single-resonance approximation of the relative density\\ of photonic states}

Another approximation that admits an analytical solution of the
evolution problem (\ref{Volterra}),~(\ref{kernel}) is a
single-resonance approximation.~Suppose that at $x\!=\!x_{r}$ the
photonic density of states $\xi(x)$ has a sharp peak of
half-width-at-half-maximum $\delta x_{r}$.~For all $x$ in the
vicinity of~$x_{r}$, the shape of $\xi(x)$ may then be roughly
approximated by the Lorentzian function of the form
\[
\xi(x)\approx{\xi(x_{r})\delta x_{r}^{2}\over{(x-x_{r})^{2}+\delta
x_{r}^{2}}}\,.
\]
The kernel (\ref{kernel}) is then easily calculated analytically
to give
\[
K(\tau-\tau^{\prime})\approx
{\hbar\Gamma(x_{r})\over{2\gamma_{0}}}{\delta x_{r}\over{2}}
\]\vspace{-0.5cm}
\[
\times{\exp[-i(x_{r}-i\delta x_{r}-x_{A})
(\tau-\tau^{\prime})-1]\over{i(x_{r}-i\delta x_{r}-x_{A})}},
\;\;\;\tau>\tau^{\prime}.
\]
Substituting this into Eq.~(\ref{Volterra}) and making the
differentiation of both sides of the resulting equation over time,
followed by the change of the integration order and one more time
differentiation, one straightforwardly arrives at a second order
ordinary homogeneous differential equation
\[
\mbox{\it\"{C}}_{u}(\tau)+i(x_{r}-i\delta x_{r}-x_{A})
\mbox{\it\.{C}}_{u}(\tau)+(\Omega/2)^{2}\,C_{u}(\tau)=0,
\]
where $\Omega=\sqrt{2\delta
x_{r}\hbar\Gamma(x_{r})/2\gamma_{0}}\,$, with the solution given
for $x_{A}\approx x_{r}$ by
\[
C_{u}(\tau)\approx\!{1\over{2}}\!\left(\!1\!+\!{\delta x_{r}
\over{\sqrt{\delta x_{r}^{2}-\Omega^{2}}}}\!\right)\!
\exp\!\!\left[-\!\left(\delta x_{r}\!-\!\sqrt{\delta x_{r}^{2}
-\Omega^{2}}\right)\!{\tau\over{2}}\right]
\]\vspace{-0.5cm}
\begin{equation}
+{1\over{2}}\!\left(\!1\!-\!{\delta x_{r}\over{\sqrt{\delta
x_{r}^{2}-\Omega^{2}}}}\!\right)\!\exp\!\!\left[-\!\left(\delta
x_{r}\!+\!\sqrt{\delta
x_{r}^{2}-\Omega^{2}}\right)\!{\tau\over{2}}\right]. \label{Cuapp}
\end{equation}
This solution is approximately valid for those atomic transition
frequencies $x_{A}$ which are located in the vicinity of the
photonic density-of-states resonances whatever the atom-field
coupling strength is.~In particular, if $(\Omega/\delta
x_{r})^{2}\!\ll\!1$, Eq.~(\ref{Cuapp})~yields the exponential
decay of the upper atomic state population probability
$|C_{u}(\tau)|^{2}$ with the rate $\hbar\Gamma(x_{r})/2\gamma_{0}$
in full agreement with Eq.~(\ref{Cumark}) obtained within the
Markovian approximation for weak atom-field coupling.~In the
opposite case, when $(\Omega/\delta x_{r})^{2}\!\gg\!1$, one has
\[
|C_{u}(\tau)|^{2}\approx e^{-\delta x_{r}\tau}
\cos^{2}\!\left(\!{\Omega\,\tau\over{2}}\!\right),
\]
and the decay of the upper atomic state population probability
proceeds via damped Rabi oscillations.~This is~the principal
signature of strong atom-field coupling which is beyond the
Markovian approximation. Expressing $\Gamma(x_{r})$ in $\Omega$ in
terms of $\xi(x_{r})$ by means of Eq.~(\ref{ksi}) and using the
approximation
$\Gamma_{0}(x)\!\approx\!\alpha^{3}2\gamma_{0}x/\hbar$ valid for
hydrogen-like atoms~\cite{Davydov}, one may conveniently rewrite
the strong atom-field coupling condition in the form
\begin{equation}
2\alpha^{3}x_{r}{\xi(x_{r})\over{\delta x_{r}}}\gg1,
\label{condstrong}
\end{equation}
from which it follows that the strong coupling regime is fostered
by high and narrow resonances in the relative density of photonic
states.

\begin{figure}[t]
\epsfxsize=8.65cm \centering{\epsfbox{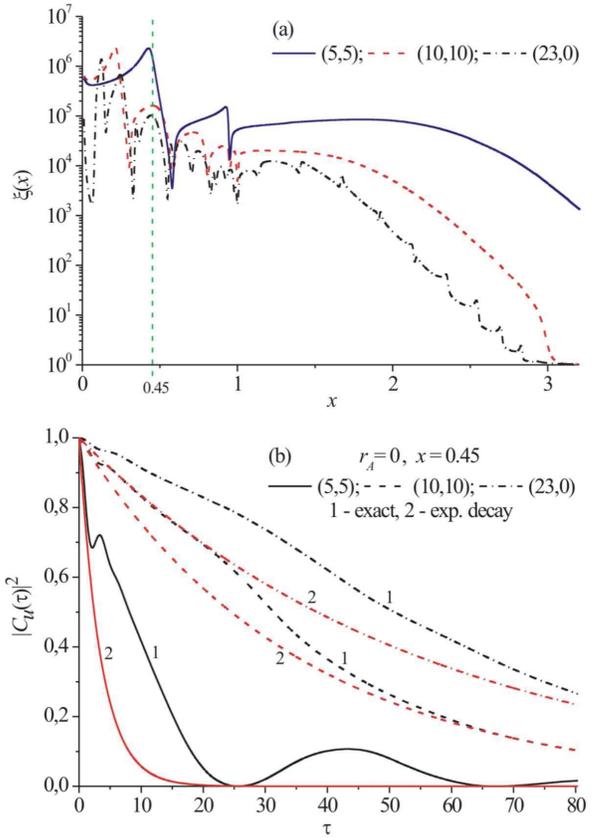}}\vskip-0.5cm
\caption{(Color online) Relative density of photonic states~(a)
and upper-level spontaneous decay dynamics (b) for the atom in the
center of different CNs.~The atomic transition frequency is
indicated by the dashed line in Fig.~\ref{fig2}~(a).} \label{fig2}
\end{figure}

\begin{figure}[t]
\epsfxsize=8.65cm \centering{\epsfbox{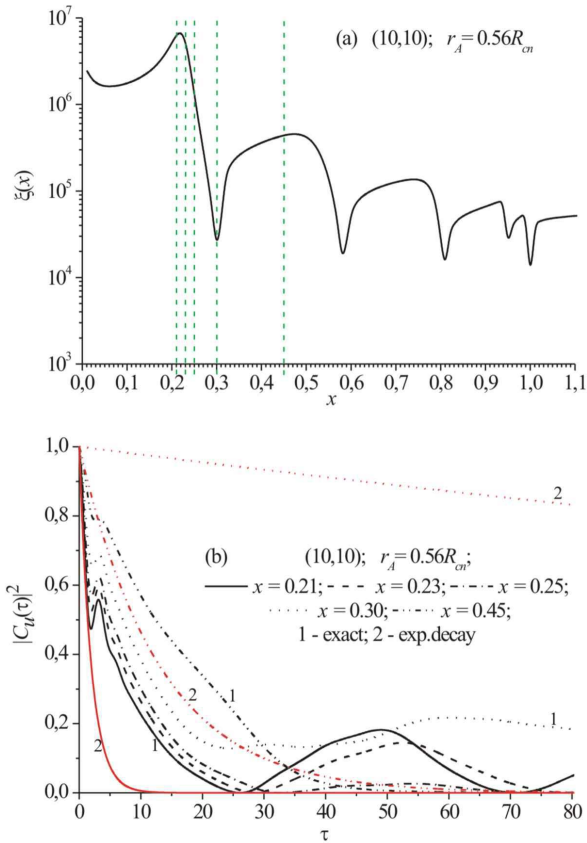}}\vskip-0.5cm
\caption{(Color online) (a)~Fragment of the relative density of
photonic states for the atom inside the (10,10) CN at distance of
3~\AA~from the wall (the situation observed experimentally for Cs
in Ref.~\cite{Jeong}).~(b)~The upper-level spontaneous decay
dynamics for different atomic transition frequencies [indicated by
the dashed lines in Fig.~\ref{fig3}~(a)] in this particular case.}
\label{fig3}
\end{figure}

\begin{figure*}[t]
\epsfxsize=17.9cm \centering{\epsfbox{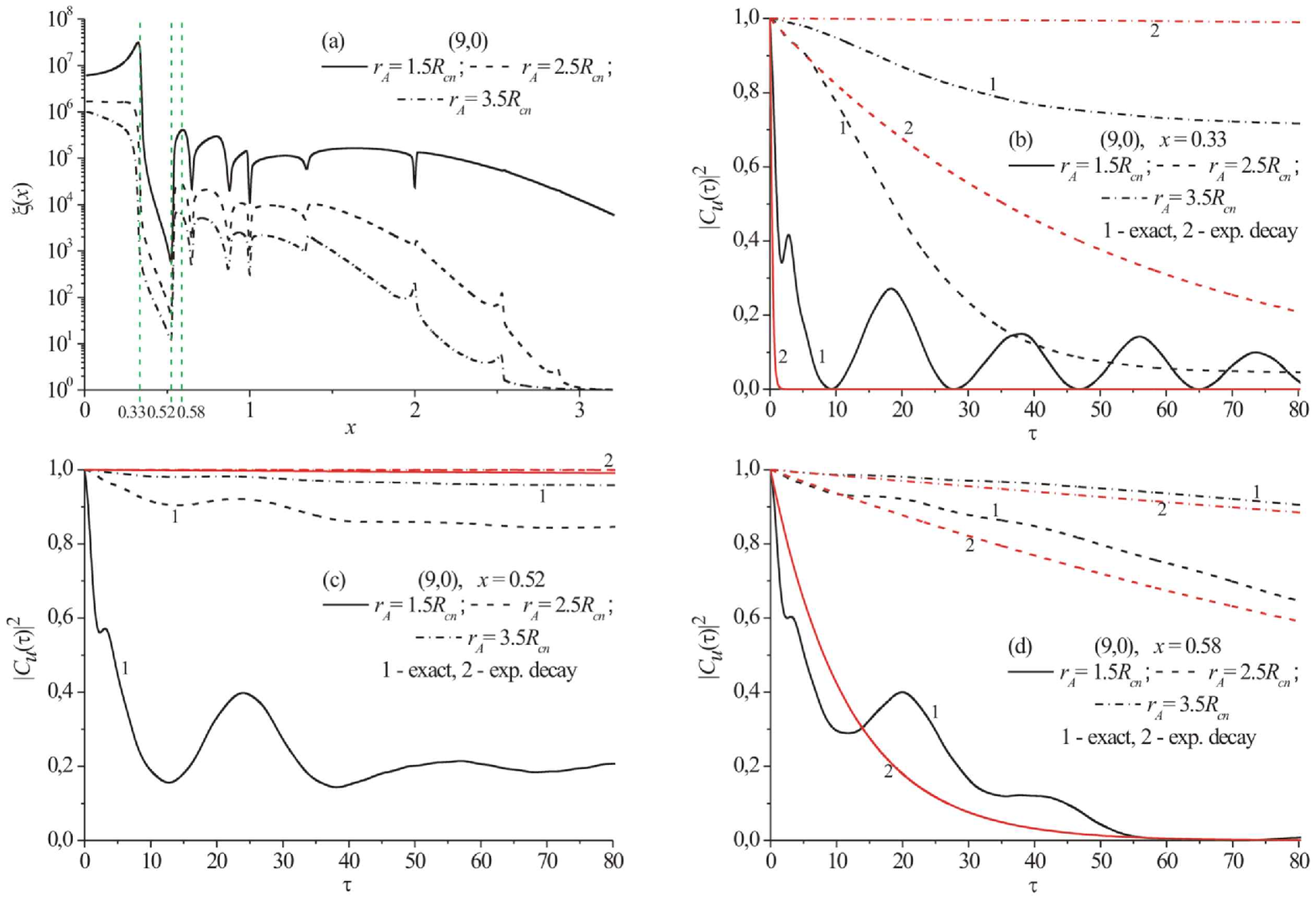}}\vskip-0.5cm
\caption{(Color online)~(a)~Relative density of photonic states
for the atom located at different distances outside the (9,0) CN.
(b, c, d)~Upper-level spontaneous decay dynamics for the three
atomic transition frequencies [indicated by the dashed lines in
Fig.~\ref{fig4}~(a)] at different atom-nanotube-surface
distances.} \label{fig4}
\end{figure*}

\section{Numerical results and discussion}\label{numerically}

To get beyond the Markovian and single-peak approximations, we
have solved Eqs.~(\ref{Volterra}) and~(\ref{kernel})
numerically.~The \emph{exact} time evolution of the upper state
population probability $|C_{u}(\tau)|^{2}$ was obtained for the
atom placed [in a way that Eq.~(\ref{condition}) was always
satisfied] in the center and near the wall inside, and at
different distances outside achiral CNs of different radii.~The
relative density of photonic states $\xi(x)$ in Eq.~(\ref{kernel})
was computed according to Eq.~(\ref{ksi}).~The CN surface axial
conductivity $\sigma_{zz}$ appearing in Eq.~(\ref{ksi}) was
calculated in the relaxation-time approximation with the
relaxation time $3\!\times\!10^{-12}$~s; the spatial dispersion of
$\sigma_{zz}$ was neglected~\cite{Slepyan}.~The free-space
spontaneous decay rate was approximated by the expression
$\Gamma_{0}(x)\!\approx\!\alpha^{3}2\gamma_{0}x/\hbar$~\cite{Davydov}.

Figure~\ref{fig2}~(a) presents $\xi(x)$ for the atom in the center
of the (5,5), (10,10) and (23,0) CNs.~It is seen to decrease with
increasing the CN radius, representing the decrease of the
atom-field coupling strength as the atom moves away from the CN
surface~\cite{Bondarev02}. To calculate $|C_{u}(\tau)|^{2}$ in
this particular case, we have fixed $x_{A}\!=\!0.45$ (indicated by
the vertical dashed line), firstly, because this transition is
located within the visible light range $0.305\!<\!x\!<\!0.574$,
secondly, because this is the approximate peak position of
$\xi(x)$ for all the three CNs.~The functions $|C_{u}(\tau)|^{2}$
calculated are shown in Figure~\ref{fig2}~(b) in comparison with
those obtained in the Markovian approximation yielding the
exponential decay.~The actual spontaneous decay dynamics is
clearly seen to be non-exponential.~For the small radius (5,5) CN,
Rabi oscillations are observed, indicating a \emph{strong}
atom-field coupling regime related to \emph{strong} non-Markovian
memory effects.~Eq.~(\ref{condstrong}) is satisfied in this
case.~With increasing the CN radius, as the value of $\xi(0.45)$
decreases, Eq.~(\ref{condstrong}) ceases to be valid and the decay
dynamics approaches the exponential one with the decay rate
enhanced by several orders of magnitude compared with that in free
space.~Note that, though the distance from the atom to the CN
surface is larger for the (23,0) CN than for the (10,10) CN, the
deviation of the actual decay dynamics from the exponential one is
larger for the (23,0) CN. This is an obvious consequence of the
influence of a small neighboring peak in the (23,0) CN photonic
density of states [Figure~\ref{fig2}~(a)].

In Ref.~\cite{Jeong}, formation of Cs-encapsulating single-wall
CNs was reported.~In a particular case of the~(10,10) CN, the
stable Cs atom/ion position was observed to be at distance of
3~\AA~from the wall.~We have simulated the spontaneous decay
dynamics for a number of typical atomic transition frequencies for
this case.~Figure~\ref{fig3}~(a) shows the density of photonic
states and the five specific transition frequencies $x_{A}$
(dashed lines) for which the functions $|C_{u}(t)|^{2}$ presented
in Figure~\ref{fig3}~(b) were calculated.~Rabi oscillations are
clearly seen to occur in the vicinity of the highest peak
($x_{A}\!\approx\!0.22$) of the photonic density of
states.~Important is that they persist for large enough detuning
values $x_{A}\!\approx\!0.21\!\div\!0.25$.~For $x_{A}\!=\!0.30$,
the density of photonic states has a dip, and the decay dynamics
exhibits no Rabi oscillations, being strongly non-exponential
nevertheless. For $x_{A}\!=\!0.45$, the intensity of the peak of
the photonic density of states is not large enough and the peak is
too broad, so that strong atom-field coupling condition
(\ref{condstrong}) is not satisfied and the decay dynamics is
close to the exponential one.

Figure~\ref{fig4}~(a) shows the density of photonic states for the
atom outside the (9,0) CN at the different distances from its
surface.~The vertical dashed lines indicate the atomic transitions
for which the functions $|C_{u}(\tau)|^{2}$ in
Figures~\ref{fig4}~(b),~(c), and (d) were calculated.~The
transitions $x_{A}\!=\!0.33$ and $0.58$ are the positions of sharp
peaks (at least for the shortest atom-surface distance), while
$x_{A}\!=\!0.52$ is the position of a dip of the function
$\xi(x)$.~Very clear underdamped Rabi oscillations are seen for
the shortest atom-surface distance at $x_{A}\!=\!0.33$
[Figure~\ref{fig4}~(b)], indicating strong atom-field coupling
with strong non-Markovity.~For $x_{A}\!=\!0.58$
[Figure~\ref{fig4}(d)], the value of $\xi(0.58)$ is not large
enough for strong atom-field coupling to occur, so that
Eq.~(\ref{condstrong}) is not fulfilled.~As a consequence, the
decay dynamics, being strongly non-exponential in general,
exhibits no clear Rabi oscillations.~For $x_{A}\!=\!0.52$
[Figure~\ref{fig4}~(c)], though $\xi(0.52)$ is comparatively
small, the spontaneous decay dynamics is still non-exponential,
approaching the exponential one only when the atom is far enough
from the CN surface.

The reason for non-exponential spontaneous decay dynamics in all
the cases considered is similar to that taking place in photonic
crystals~\cite{John}.~When the atom is closed enough to the CN
surface, an absolute value of the relative density of photonic
states is large and its frequency variation in the neighborhood of
a specific atomic transition frequency essentially influences the
time behavior of the kernel~(\ref{kernel}) of evolution
equation~(\ref{Volterra}). Physically, this means that the
correlation time of the electromagnetic vacuum is not negligible
on the time scale of the evolution of the atomic system, so that
atomic motion memory effects are important and the Markovian
approximation in the kernel~(\ref{kernel}) is inapplicable.

\section{Conclusions}\label{conclusion}

The effects we predict will yield an additional structure in
optical absorbance/reflectance spectra (see,
e.g.,~\cite{Li,Marinop}) of atomically doped CNs in the vicinity
of the energy of an atomic transition. Weak non-Markovity of the
spontaneous decay (non-exponential decay with no Rabi
oscillations) will cause an asymmetry of an optical spectral
line-shape (similar to exciton optical absorbtion line-shape in
quantum dots~\cite{Bondarev03}).~Strong non-Markovity of the
spontaneous decay (non-exponential decay with fast Rabi
oscillations) originates from strong atom-vacuum-field coupling
with the upper state of the system splitted into two "dressed"
states.~This will yield a two-component structure of optical
spectra similar to that observed for excitonic and intersubband
electronic transitions in semiconductor quantum
microcavities~\cite{Weisbuch1,Sorba}.

Summarizing, we have developed the quantum theory of the
spontaneous decay of an excited two-level atom near a carbon
nanotube.~In describing atom-field interaction, we followed an
electromagnetic field quantization scheme developed for dispersing
and absorbing media in Refs.~\cite{Dung,Welsch}.~This quantization
formalism was adopted by us for a particular case of an~atom near
an infinitely long single-wall CN.~We derived the evolution
equation of the upper state of the system~and, by solving~it
numerically, demonstrated a~strictly non-exponential spontaneous
decay dynamics in the case where the atom is close enough to the
CN surface.~In certain cases, namely when the atom is close enough
to the nanotube surface and the atomic transition frequency is in
the vicinity of the resonance of the photonic density of states,
the system exhibits vacuum-field Rabi oscillations, a principal
signature of strong atom-vacuum-field coupling, --- a result
already detected for quasi-2D excitonic and intersubband
electronic transitions in semiconductor quantum
microcavities~\cite{Weisbuch1,Sorba} and never reported for
atomically doped CNs so far.~This is the result of strong
non-Markovian memory effects arising from the rapid frequency
variation of the photonic density of states near the nanotube.~The
non-exponential decay dynamics gives place to the exponential one
if~the atom moves away from the CN surface.~Thus, the
atom-vacuum-field coupling strength and the character of the
spontaneous decay dynamics, respectively, may be controlled~by
changing the distance between the atom and CN surface by means of
a proper preparation of atomically doped CNs.~This opens routes
for new challenging nanophotonics applications of atomically doped
CN systems as various sources of coherent light emitted by dopant
atoms.

Finally, we would like to emphasize a general character of the
results we obtained.~We have shown that similar to semiconductor
microcavities~\cite{Sorba} and photonic band-gap
materials~\cite{John}, carbon nanotubes may qualitatively change
the character of atom-electromagnetic-field interaction, yielding
strong atom-field coupling --- an important phenomenon necessary,
e.g., for quantum information
processing~\cite{Raimond,Abstreiter,Steel}.~The present paper
dealt with the simplest manifestation of this general phenomenon
--- vacuum-field Rabi oscillations in the atomic spontaneous decay
dynamics near a single-wall carbon nanotube.~However, similar
manifestations of strong atom-field coupling may occur in many
other atom-electromagnetic-field interaction processes in the
presence of CNs, such as, e.g., dipole-dipole interaction between
atoms by means of a~vacuum photon exchange~\cite{Agarwal}, or
cascade spontaneous transitions in three-level atomic
systems~\cite{Dalton}.

\acknowledgments

We gratefully acknowledge numerous discussions with Dr.
G.Ya.~Slepyan and Prof. I.D.~Feranchuk. I.B.~thanks the Belgian
OSTC. The work was performed within the framework of the Belgian
PAI-P5/01 project.

\appendix

\section{Green tensor of a~single-wall carbon nanotube}\label{appA}

The form of the classical electromagnetic field Green tensor in
Eqs.~(\ref{Erw}) and (\ref{GreenequCN}) depends on the presence of
external radiative sources (an atom in our case) and their
position (inside or outside) with respect to a~nanotube.~Since the
problem has a cylindric symmetry, we assign the orthonormal
cylindric basis
$\{\mathbf{e}_{r},\mathbf{e}_{\varphi},\mathbf{e}_{z}\}$ (see
Figure~\ref{fig1}) in such a way that $\mathbf{e}_{z}$ is directed
along the nanotube axis and
$\mathbf{r}_{A}=r_{A}\mathbf{e}_{r}=\{r_{A},0,0\}$.

We use the representation
\begin{equation}
G_{\alpha z}(\mathbf{r},\mathbf{r}_{A},\omega)=
\left({1\over{k^{2}}}\nabla_{\alpha}\nabla_{z}+ \delta_{\alpha
z}\right)g(\mathbf{r},\mathbf{r}_{A},\omega), \label{Greenrepres}
\end{equation}
where $\alpha=\{r,\varphi,z\}$ and the function
$g(\mathbf{r},\mathbf{r}_{A},\omega)$ is the Green function of the
Helmholtz equation.~Substituting Eq.~(\ref{Greenrepres}) into
Eq.~(\ref{GreenequCN}), one straightforwardly obtains
\begin{equation}
(\Delta+k^{2})\,g(\mathbf{r},\mathbf{r}_{A},\omega)=-\delta(\mathbf{r}-\mathbf{r}_{A})
\label{pointsource}
\end{equation}
with a known solution
\begin{equation}
g_{0}(\mathbf{r},\mathbf{r}_{A},\omega)={1\over{4\pi}}\,
{e^{ik|\mathbf{r}-\mathbf{r}_{A}|}\over{|\mathbf{r}-\mathbf{r}_{A}|}}
\label{g0}
\end{equation}
satisfying the radiation condition at infinity (see, e.g.,
\cite{Davydov}).~In our case, however, the functions $G_{\alpha
z}(\mathbf{r},\mathbf{r}_{A},\omega)$ and
$g(\mathbf{r},\mathbf{r}_{A},\omega)$ are imposed one more set of
boundary conditions.~They are the boundary conditions on the
surface of the CN. Using simple relations (valid for
$\mathbf{r}\ne\mathbf{r}_{A}$ under the Coulomb-gauge condition)
\begin{equation}
E_{\alpha}(\mathbf{r},\omega)=ikG_{\alpha
z}(\mathbf{r},\mathbf{r}_{A},\omega) \label{Ealpha}
\end{equation}
and
\begin{equation}
H_{\alpha}(\mathbf{r},\omega)=-{i\over{k}}\!\!\sum_{\beta,\gamma=r,\varphi,z}
\!\!\!\epsilon_{\alpha\beta\gamma}\nabla_{\beta}E_{\gamma}(\mathbf{r},\omega)
\label{Halpha}
\end{equation}
($\epsilon_{\alpha\beta\gamma}$ is the totally antisymmetric unit
tensor of rank 3), they can be derived from the classical
electromagnetic field boundary conditions of the form
\begin{equation}
\underline{E\!}_{\,\varphi}|_{r=R_{cn}+0}-
\underline{E\!}_{\,\varphi}|_{r=R_{cn}-0}=0\,,
\label{boundaryEphiCN}
\end{equation}\vspace{-0.5cm}
\begin{equation}
\underline{E\!}_{\,z}|_{r=R_{cn}+0}-
\underline{E\!}_{\,z}|_{r=R_{cn}-0}=0\,, \label{boundaryEzCN}
\end{equation}\vspace{-0.5cm}
\begin{equation}
\underline{H\!}_{\,\varphi}|_{r=R_{cn}+0}-
\underline{H\!}_{\,\varphi}|_{r=R_{cn}-0}
={4\pi\over{c}}\,\sigma_{zz}(R_{cn},\omega)
{\underline{E\!}_{\,z}}|_{r=R_{cn}}, \label{boundaryHphiCN}
\end{equation}\vspace{-0.5cm}
\begin{equation}
\underline{H\!}_{\,z}|_{r=R_{cn}+0}-
\underline{H\!}_{\,z}|_{r=R_{cn}-0}=0 \label{boundaryHzCN}
\end{equation}
(spatial dispersion neglected) obtained in Ref.~\cite{Slepyan}.

Let $r_{A}\!>\!R_{cn}$ (the atom outside the CN) to be specific.
Then, $g(\mathbf{r},\mathbf{r}_{A},\omega)$ may be represented in
the form
\begin{equation}
g(\mathbf{r},\mathbf{r}_{A},\omega)=\left\{\begin{array}{ll}
g_{0}(\mathbf{r},\mathbf{r}_{A},\omega)+g^{(+)}(\mathbf{r},\omega)\,,&r>R_{cn}\\[0.1cm]
g^{(-)}(\mathbf{r},\omega)\,,&r<R_{cn}\end{array}\right.
\label{gdef}
\end{equation}
where $g_{0}(\mathbf{r},\mathbf{r}_{A},\omega)$ is the point
radiative atomic source function defined in Eq.~(\ref{g0}) and
$g^{(\pm)}(\mathbf{r},\omega)$ are unknown nonsingular functions
satisfying the homogeneous Helmholtz equation and the radiation
conditions at infinity.~We seek them using integral decompositions
over the modified cylindric Bessel functions $I_{p}\,$ and $K_{p}$
as follows~\cite{Jackson}
\begin{equation}
g^{(\pm)}(\mathbf{r},\omega)=\!\!\sum_{p=-\infty}^{\infty}\!\!e^{ip\varphi}\!\!
\int_{C}\left\{\!\!\begin{array}{l}A_{p}(h)\,K_{p}(vr)\\[0.1cm]
B_{p}(h)\,I_{p}(vr)\end{array} \!\!\right\}e^{ihz}dh
\label{gpmexpan}
\end{equation}
and
\[
g_{0}(\mathbf{r},\mathbf{r}_{A},\omega)=
{1\over{(2\pi)^{2}}}\sum_{p=-\infty}^{\infty}\!\!e^{ip\varphi}\!\!
\]\vspace{-0.5cm}
\begin{equation}
\times\int_{C}I_{p}(vr)\,K_{p}(vr_{A})\,e^{ihz}dh\,,
\;\;\;r_{A}\geq r\,, \label{g0expan}
\end{equation}
where $A_{p}(h)$ and $B_{p}(h)$ are unknown functions to be found
from the boundary conditions
(\ref{boundaryEphiCN})--(\ref{boundaryHzCN}) in view of
Eqs.~(\ref{Greenrepres}),~(\ref{Ealpha}) and (\ref{Halpha}),
$v\!=\!v(h,\omega)\!=\!\sqrt{h^{2}-k^{2}}$. The integration
contour $C$ goes along the real axis of the complex plane and
envelopes the branch points $\pm k\,$ from below and from above,
respectively.

The boundary conditions
(\ref{boundaryEphiCN})--(\ref{boundaryHzCN}) with
Eqs.~(\ref{Greenrepres}), (\ref{Ealpha}) and (\ref{Halpha}) taken
into account yield the following two independent equations for the
scalar Green function~(\ref{gdef})
\[
g_{0}(\mathbf{r},\mathbf{r}_{A},\omega)|_{r=R_{cn}}+
g^{(+)}(\mathbf{r},\omega)|_{r=R_{cn}}
=g^{(-)}(\mathbf{r},\omega)|_{r=R_{cn}},
\]
\[
{\partial g^{(+)}(\mathbf{r},\omega)\over{\partial
r}}|_{r=R_{cn}}-{\partial g^{(-)}(\mathbf{r},\omega)\over{\partial
r}}|_{r=R_{cn}}+
\]\vspace{-0.5cm}
\[
\beta(\omega)\!\left(\!{\partial^{2}\over{\partial
z^{2}}}\!+\!k^{2}\!\right)\!g^{(-)}(\mathbf{r},\omega)|_{r=R_{cn}}\!\!=\!-{\partial
g_{0}(\mathbf{r},\mathbf{r}_{A},\omega)\over{\partial
r}}|_{r=R_{cn}},
\]
where $\beta(\omega)\!=\!4\pi
i\,\sigma_{zz}(R_{cn},\omega)/\omega$.~Substituting the integral
decompositions (\ref{gpmexpan}) and (\ref{g0expan}) into these
equations, one obtains the set of two simultaneous algebraic
equations for the functions $A_{p}(h)$ and $B_{p}(h)$.~The
function $A_{p}(h)$ we need (we only need the Green function in
the region where the atom is located) is found by solving this set
with the use of basic properties of cylindric Bessel functions
(see, e.g.,~\cite{Abramovitz}). In so doing, one has
\[
A_{p}(h)=-{R_{cn}\beta(\omega)\,v^{2}I_{p}^{2}(vR_{cn})
K_{p}(vr_{A})\over{(2\pi)^{2}[1+\beta(\omega)\,v^{2}R_{cn}
I_{p}(vR_{cn})K_{p}(vR_{cn})]}}\,.
\]
These $A_{p}(h)$, being substituted into Eq.~(\ref{gpmexpan}),
yield the function $g^{(+)}(\mathbf{r},\omega)$ sought.~The latter
one, in view of Eq.~(\ref{gdef}), results in the scalar
electromagnetic field Green function of the form
\[
g(\mathbf{r},\mathbf{r}_{A},\omega)=g_{0}(\mathbf{r},\mathbf{r}_{A},\omega)
-{R_{cn}\over{(2\pi)^{2}}}
\]\vspace{-0.5cm}
\begin{equation}
\times\!\!\!\sum_{p=-\infty}^{\infty}\!\!
e^{ip\varphi}\!\!\int_{C}{\beta(\omega)\,v^{2}I_{p}^{2}(vR_{cn})
K_{p}(vr_{A})K_{p}(vr)\over{1+\beta(\omega)\,v^{2}R_{cn}
I_{p}(vR_{cn})K_{p}(vR_{cn})}}\;e^{ihz}dh\,, \label{g}
\end{equation}
where $r_{A}\!\ge\!r\!>\!R_{cn}$.~One may show in a similar way
that the function $g(\mathbf{r},\mathbf{r}_{A},\omega)$ for
$r\!\le\!r_{A}\!<\!R_{cn}$ is obtained from Eq.~(\ref{g}) by means
of a simple symbol replacement $I_{p}\!\leftrightarrow\!K_{p}$ in
the numerator of the integrand.

Knowing $g(\mathbf{r},\mathbf{r}_{A},\omega)$, one may easily
compute the components of the electromagnetic field Green tensor
$G_{\alpha z}(\mathbf{r},\mathbf{r}_{A},\omega)$ according to
Eq.~(\ref{Greenrepres}).

\section{Density of photonic states near a single-wall carbon
nanotube}\label{appB}

The relative density of photonic states in an atomic spontaneous
decay process near a CN is defined by the ratio [see
Eqs.~(\ref{dimless})--(\ref{ksi})]
\begin{equation}
\xi(\omega)={\mbox{Im}\,G_{zz}(\mathbf{r}_{A},\mathbf{r}_{A},\omega)
\over{\mbox{Im}\,G^{v}_{zz}(\mathbf{r}_{A},\mathbf{r}_{A},\omega)}}\,,
\label{relphdens}
\end{equation}
where
\begin{equation}
G_{zz}(\mathbf{r}_{A},\mathbf{r}_{A},\omega)=
\left({1\over{k^{2}}}{\partial^{2}\over{\partial z^{2}}}+1\right)
g(\mathbf{r},\mathbf{r}_{A},\omega)|_{\mathbf{r}=\mathbf{r}_{A}}
\label{Gzz}
\end{equation}
according to Eq.~(\ref{Greenrepres}),
$g(\mathbf{r},\mathbf{r}_{A},\omega)$ is given by Eq.~(\ref{g})
for $r_{A}\!\ge\!r\!>\!R_{cn}$ and by the same equation with the
symbol replacement $I_{p}\!\leftrightarrow\!K_{p}$ in the
numerator of the integrand for $r\!\le\!r_{A}\!<\!R_{cn}$.

Substituting Eq.~(\ref{g}) into Eq.~(\ref{Gzz}), making
differentiation and passing to the limit
$\mathbf{r}\!\rightarrow\!\mathbf{r}_{A}$, followed by the
substitution of the result into Eq.~(\ref{relphdens}), one has for
$r_{A}\!>\!R_{cn}$
\begin{equation}
\xi(\omega)=1+{3R_{cn}\over{2\pi k^{3}}} \label{phdens}
\end{equation}\vspace{-0.5cm}
\[
\times\,\mbox{Im}\!\!\!\sum_{p=-\infty}^{\infty}\int_{C}{\beta(\omega)\,v^{4}
I_{p}^{2}(vR_{cn})K_{p}^{2}(vr_{A})\over{1+R_{cn}\,\beta(\omega)\,v^{2}
I_{p}(vR_{cn})K_{p}(vR_{cn})}}\;dh\,.
\]
For $r_{A}\!<\!R_{cn}$, Eq.~(\ref{phdens}) is modified either by
the symbol replacement $I_{p}\!\leftrightarrow\!K_{p}$ or by the
Bessel function argument replacement
$r_{A}\!\leftrightarrow\!R_{cn}$ in the numerator of the
integrand. In dimensionless variables (\ref{dimless}),
Eq.~(\ref{phdens}) is rewritten to give Eq.~(\ref{ksi}).

\end{document}